# Dimensional analysis on superconducting plasmonics

T. Dogan[1]

[1] The Department of Applied Physics, Eindhoven University of Technology, 5600 Eindhoven, The Netherlands (e-mail: t.dogan@tue.nl).

**Abstract**

Plasmonic response of superconductors at various dimensions are addressed in this paper. All possible parameter space is discussed and considered for theoretical demonstration towards possible future experiments. The most critical parameters in each dimensional state are found to be carrier concentration (3D), wavevector (2D), surrounding permittivity (1D) and particle radius (0D). However, it is also demonstrated that unless all parameters combined with optimal conditions, the superconducting plasmonic system does not result in an ideal lossless environment. Furthermore, under all conditions, lower temperature decreases the loss content affiliated with electron and phonon scattering.

## Introduction

Plasmonic devices have been increasingly attracting attention to be used with variety of materials such as dielectric waveguides, organic solar cells, metal-based biosensors and high transmission superconductors [1-5]. The fundamental phenomenon behind many of the plasmonic applications is surface plasmon polariton (SPP) which is mostly investigated on metal-dielectric interfaces. Further development and use of SPP, however, is hindered by high generation of losses in metals due to electron-electron interactions, interface roughness and phonon scattering [6]. A recent study has demonstrated that even if those loss mechanisms were to be minimized, the field confinement or the Landau damping would still limit the performance of SPP for nanometer scale dimensions [6]. In order to achieve lossless plasmonic systems, superconductors have emerged as an alternative material set [7]. Although the first theoretical and experimental investigations of surface/plasmon waves on superconductors were performed more than two decades ago [8-9], the application-wise interest has only lately increased after evolution of high temperature superconductors. Buisson et. al. was the first to demonstrate that under optimal conditions, propagating plasma modes live on thin superconducting films [9]. Tian et. al. and Tsiatmas et. al. produced enhanced THz transmission using hole arrays on bulk superconductors [10-11]. Thomas et. al. has experimentally proved the possibility of using auxiliary coupling centers that coupled light to otherwise difficult superconducting species [12]. There are many exemplary studies on superconductors with various dimensions, and in this paper material parameters and environmental factors are discussed to provide a unified dimensional approach to superconducting plasmonics [13-16].

Considering the two-fluid model, below a critical temperature ($T_c$), superconductors are composed of two types of carriers, normal electrons and cooper pairs [17]. While the amount of the former decreases as temperature approaches to zero, the latter reaches its maximum value at absolute zero. According to well-known Casimir-Gorter relation, both carrier concentrations are dependent on temperature as follows

$$n_s = n_0[1 - (t)^4] \quad , \quad n_n = n_0(t)^4 \qquad (1)$$

where $t = T/T_c$ is the scaled temperature with $T$ ambient temperature and $T_c$ superconducting critical temperature, $n_0$ is the total density of electrons, $n_s$ and $n_n$ are super and normal electron densities. Even though the critical temperature is material dependent, the scaled temperature parameter $t$ will be considered material independent variable between 0 and 1. This two-fluid model relation is credible for all superconductors in any shape/dimension. Furthermore, since SPP is the collective oscillation of electrons in the presence of momentum matched electromagnetic field, the distribution of electrons among normal and super state influences the plasmonic behavior of the superconductor greatly. The plasma frequency of electrons in each respective state is given by [18]

$$\omega_i = \sqrt{\frac{n_i e}{m\varepsilon_0}} \frac{\kappa^{(3-d)/2}}{\epsilon} \qquad (2)$$

Here, $\omega_i$ and $n_i$ denotes the plasma frequency and density of electrons in $i = n, s$ state. $e$ is elementary charge, $m$ is electron mass, $\varepsilon_0$ is vacuum permittivity. It should be noted that the mass of electrons differs slightly for cooper pairs (in the order of $10^{-6}$ $m_0$), and this deviation is omitted here since it is considered negligible [20]. The plasma frequency equation is exactly of the same form as for metals. The dispersion relation for superconductors also resemble to that of metals except for inclusion of cooper pairs. An important difference from metals is that the frequency of the SPP must be sufficiently smaller than the energy gap frequency of the superconductors because the lossless nature of the superconducting electrons will be destroyed near that frequency [16]. The energy gap of a superconductor can be defined as $\Delta_g = \alpha k_b T_c$, and the corresponding frequency is $\omega_g = 4\pi\Delta_g/\hbar$ where $k_b$ is Boltzmann constant, $\hbar$ is Planck constant. The value of $\alpha$ ranges between 1.75 and 2 depending on the superconductor. In order to capture any superconductor, it is safe to assume that the gap frequency is in the order of THz, and in this study the maximum operation frequency will be fixed at $\omega = 1$ THz.

In the rest of this paper, each dimensional analysis is carried out separately, and vital parameters for utilization of plasma modes on them are specified.

## Surface Plasmons on 3D Superconductors

As the name stresses surface plasmon polaritons exist on the surface. However, their behavior depends on whether the charges are confined to the surface or not because SPPs are evanescent waves that decay along the direction normal to the surface [20]. In a 3D system the electrons are distributed within the whole material. As in metals, the dispersion relation in superconductors also relies on the frequency dependent dielectric functions.

$$\beta = \beta' + i\beta'' = \frac{\omega}{c}\sqrt{\frac{\varepsilon_s' \varepsilon_d}{\varepsilon_s' + \varepsilon_d}} + i\frac{\omega}{c}\left[\frac{\varepsilon_s''}{2(\varepsilon_s')^2}\left(\frac{\varepsilon_s' \varepsilon_d}{\varepsilon_s' + \varepsilon_d}\right)^{\frac{3}{2}}\right] \quad (3)$$

$$\varepsilon_s = \varepsilon_s' + i\varepsilon_s'' = 1 - \frac{\omega_s^2}{\omega^2} - \frac{\omega_n^2}{\omega^2 + \gamma^2} + i\frac{\omega_n^2 \gamma}{\omega(\omega^2 + \gamma^2)} \quad (4)$$

where $\beta'$ and $\beta''$ define the real and imaginary parts of the propagation constant, $\gamma$ is scattering rate and $\varepsilon_d$ is the permittivity of the ambient. The dielectric function also has real and imaginary parts denoted by $\varepsilon_s'$ and $\varepsilon_s''$. The physical meaning behind $\beta''$ is the energy loss of the SPP during propagation. A generic term of propagation length, at which distance the energy decreases by one exponential order is given by $L = 1/2\beta''$. As suggested by Khurgin et. al., for 3D plasmonic system a figure of merit comparing $\beta'$ and $\beta''$ would then be $FOM = \beta'/2\beta''$ [6].

When the super and normal plasma frequencies are considered, the dielectric function ($\varepsilon_s$) appears to depend on four variables $n_0, \gamma, \omega, t$. While the total electron density and scattering rate are material properties, the operation frequency and temperature are environmental factors. Furthermore, the dispersion relation ($\beta$) also depends on the dielectric constant of the ambient ($\varepsilon_d$). All of the parameters can be adjusted to obtain optimal plasmonic conditions. Naturally depending on the type of application, the best parameter set might vary since frequency is an important factor.

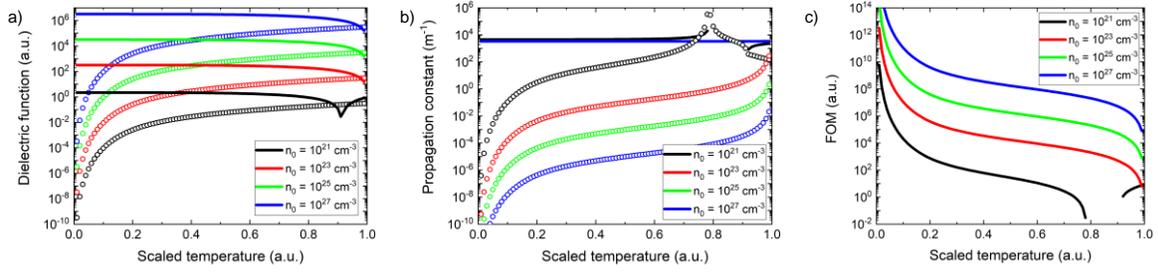

**Figure 1**. (a) Dielectric function of 3D superconductors for varying carrier concentration and temperature. The frequency is fixed to 1THz. (b) Real and imaginary parts of the propagation constant of 3D superconductors for varying carrier concentration and temperature. (c) Figure of merit results of 3D superconductors calculated from (b).

Among both conventional and unconventional superconductors the lowest bulk carrier concentration is around $n_0 = 10^{21}\ cm^{-3}$ (for example in STO $n_0 = 4 \times 10^{21}\ cm^{-3}$), while most of the high Tc superconductors have $n_0 = 10^{28}\ cm^{-3}$. Doping of the superconductors is also possible in diverse manners [17]. The scattering rate also varies among materials and depends on temperature. Albeit certain variations, the rate is given by the simplified equation $\gamma = \xi t^{1.5}$, here $\xi = 10^{13} s^{-1}$ will be assumed for all materials [21]. As mentioned above, the frequency can reach a maximum of $10^{12}$ Hz. Lastly, the superconductor could have an interface with any dielectric material, but only vacuum will be investigated here ($\varepsilon_d = 1$).

The direct effect of carrier concentration and temperature on dielectric function is demonstrated in Fig. 1a. While both real and imaginary parts have strong dependence on the carrier concentration, only the imaginary part approaches zero at very low temperatures. Additionally, in order for plasmonic waves to bound to the surface the real part must be relatively close to the permittivity of the ambient. In the case of vacuum or air ($\varepsilon_d \sim 1$), $|\varepsilon'_s|$ should stay below 100. Considering this fact, only superconductors that have carrier concentration below $10^{23}$ cm$^{-3}$ are eligible to support surface plasmons. Additionally, the low end ($10^{21}$) of the carrier concentration spectrum also shows that the real part becomes infinitesimal at a certain temperature. In Fig. 1a, the drop to zero occurs due to the superposition among normal and super carrier concentrations that leads to diminishing real dielectric function. At that exact point, the superconductor becomes a perfect absorber. According to these results, the best approach would be to fine adjust $\varepsilon'_s$ for best SPP through doping the superconductor and decrease the temperature for very low $\varepsilon''_s$. Such a setting would practically create a state-of-the-art plasmonic behavior. However, it should be strongly stressed that these results are for 1 THz and lowering of the frequency also requires lowering of the carrier concentration as can be seen from Eqn. 4.

Fig. 1b shows the corresponding results of propagation constant with respect to carrier concentration and temperature. The variation of $\beta'$ is much smaller when compared to $\beta''$ for different temperature as latter depends strongly on $\varepsilon''_s$. For low concentrations the $\beta'$ values depart from the wavevector of light at corresponding frequency ($k = \omega/c$), which ensures proper bounding of the wave to the surface. In Fig. 1c, a more elaborate result of FOM is given. For comparison, an ideal metal surface plasmon has a FOM around 1000, this is also drawn in Fig. 1c. It can be understood that very low carrier density is not in itself enough for best performance but a combination of all factors $n_0, \gamma, \omega, t, \varepsilon_d$ needs to be taken into account. However, superconductors ensure that at low temperatures plasmonic parameters boost up due to diminishing loss elements.

## Surface Plasmons on 2D Superconductors

For the two-dimensional case, most of the above equations still hold except for the plasma frequency. In this case, there also appears a direct dependence on the square-root of the wavevector ($\kappa$) and the permittivity of the upper ($\varepsilon_u$) and lower ($\varepsilon_l$) bases [9,15]. The simplified version of the equation is as follows

$$\omega_i = \sqrt{\frac{n_i e}{m \varepsilon_0}} \sqrt{\frac{\kappa}{(\varepsilon_u + \varepsilon_l)}} \quad (5)$$

Here, $\kappa$ is inversely proportional to physical length ($L$) of the superconductor, $\kappa = \pi/L$. Additionally, it should be noted that various modes are also available at integer multiplies of the wavevector. However, only the first mode is examined in this study since the other modes might be very weak to observe experimentally. Above relation between plasma frequency and the wavevector plays a very crucial role due to the occurring tradeoff between the operation frequency and the length. High frequencies around 1 THz are only applicable to 2D superconductors with micrometer lengths. However, in order for signals to propagate longer, the frequency must be lowered by at least 2-3 orders of magnitude. For example, Buisson et. al., have demonstrated surface waves of GHz frequencies on 2D Aluminum superconductor with lengths around millimeter range [9].

In terms of carrier concentrations, the 2D superconductors have a variety of possibilities. For example, an interface between Lanthanum Aluminate-Strontium Titanate (LAO-STO) has a two-dimensional electron gas (2DEG) that becomes superconducting at very low temperatures (around 200 mK) [17]. The carrier concentration for this instance is around $10^{13}$ cm$^{-2}$. For most metals, on the other hand, the density can be as high as $10^{18}$ cm$^{-2}$. While 2DEG is already limited to 2D, it is of central importance to achieve such regularity in thin films of bulk materials. Therefore, the thickness of a superconductor must be in the range of coherence length for maximum reliability as a 2D object [9].

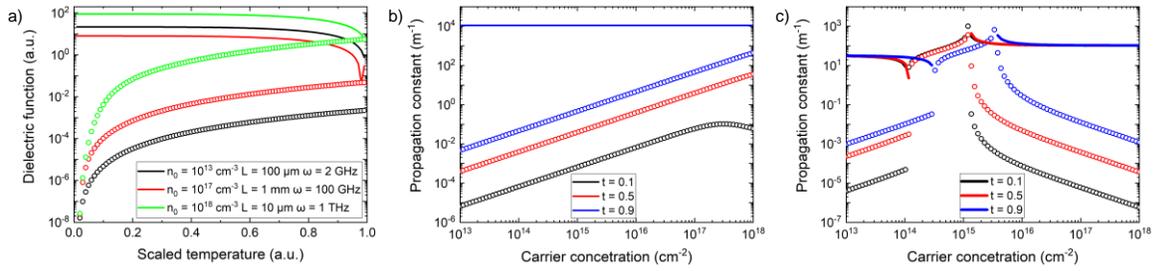

**Figure 2.** (a) Dielectric function of 2D superconductors for selected combinations of carrier concentration, superconductor length and frequency plotted against changing temperature. (b) Real and imaginary parts of the propagation constant of 2D superconductors for varying carrier concentration at $\omega = 10\ GHz$ and $L = 1\ mm$. (c) Real and imaginary parts of the propagation constant of 2D superconductors for varying carrier concentration at $\varepsilon'_s = 100$.

Another essential factor is the permittivity of the base since 2D superconductor has to lie on a substrate. A normal glass or sapphire base has permittivity in the order of 10, but at low temperatures the permittivity of STO could reach values in the order of 1000 [9]. This has a direct influence on the generation of the surface waves, and it should be considered together with the choice of superconductor and its carrier concentration. For example, a low permittivity insulator would work better for low carrier concentration superconductor.

In Fig. 2a, the dielectric function vs temperature is plotted for various combinations of carrier concentration, frequency and the length. As the limiting case, the propagation of THz frequency seems only possible for maximum lengths on the order of 10 um if the carrier concentration is around $10^{18}$ cm$^{-2}$. The substrate is considered to have a permittivity of 10. However, surface waves can also exist for a longer range (mm or cm) at lower signal frequencies.

Fig. 2a only plots some of the exemplary combinations of parameters that could give rise to observable SPP waves on a 2D superconductor.

The change in propagation can be more clearly understood when $\beta'$ and $\beta''$ is plotted for different carrier concentrations. In Fig. 2b, this plot is given for temperatures $t = 0.1, 0.5, 0.9$. The frequency and the length are set to 10 GHz and 1 mm for these plots. An interesting fact is that there occur non-propagating plasma modes at certain concentrations (between $10^{14} - 10^{15}$ cm$^{-2}$) for the given parameters. It should be noted that the location of this lossy part can be adjusted for higher or lower carrier concentrations by changing operation frequency and/or the length. Furthermore, $\beta''$ follows a symmetric trend based on the central peak, which allows to adjust the behavior for the best plasmonic conditions. For example, if the loss is to be minimized at high concentrations the frequency can be decreased and the whole graph shifts towards lower concentration enabling disproportion between $n_0$ and $\beta''$. Additionally, the temperature is only effective for the loss mechanism, meaning establishing strong decrease in losses does not affect propagation constant.

Considering the behavior of dielectric function with respect to carrier concentration, it becomes obvious from Fig. 2a that frequency and wavevector must scale with the density in order to accompany appropriately bound surface waves. Fig. 2c demonstrates the change of $\beta''$ with carrier concentration, and the other parameters are arranged such that the real part of the dielectric function is fixed to -100. There is a definite reduce in loss not only with temperature but also with carrier concentration. Moreover, decreased density also enables much longer survival of plasmonic signals.

## Surface Plasmons on quasi-1D Superconductors

The plasma mode on quasi-1D superconductors was first suggested by Mooij and Schön [14]. In this case, the bulk superconductor is shrunk in transverse dimensions with a diameter longer than the coherence length. However, the plasma mode could only survive up to a radius of around 50 nm. The plasma frequencies are defined in a similar manner as in 3D (Eqn 1). However, due to existence of normal electrons, there occurs damping with the frequency

$$\omega_d = \frac{\omega_s^2 \varepsilon_0}{\sigma_n t^4} \tag{6}$$

Here, $\sigma_n$ defines the normal state conductivity. Additionally, the scattering rate for quasi-1D mode has a linear dependence on the wavevector.

$$\gamma = c_p \kappa \tag{7}$$

Definitions for the plasma speed $c_p$ and the wavevector are given below

$$c_p^2 = \phi \omega_s^2 A \tag{8}$$

$$\kappa = \frac{\exp(-\delta \varepsilon_d)}{r} \tag{9}$$

Where $r$ and $A$ are the cross-section area and radius of the superconducting wire. $\phi$ and $\Delta$ are material dependent constants. Although these constants are defined in terms of capacitance per unit length, further calculations are omitted since generic values are sufficient for understanding purposes. Considering the damping frequency and the scattering rate explained above, the dielectric function is represented by

$$\varepsilon_s = \varepsilon_s' + i\varepsilon_s'' = 1 - \frac{\omega_d^2}{\frac{\omega^2 \omega_d^2}{\gamma^2} + \gamma^2} + i \frac{\omega_d}{\omega \left(\frac{\omega^2 \omega_d^2}{\gamma^4} + 1\right)} \tag{10}$$

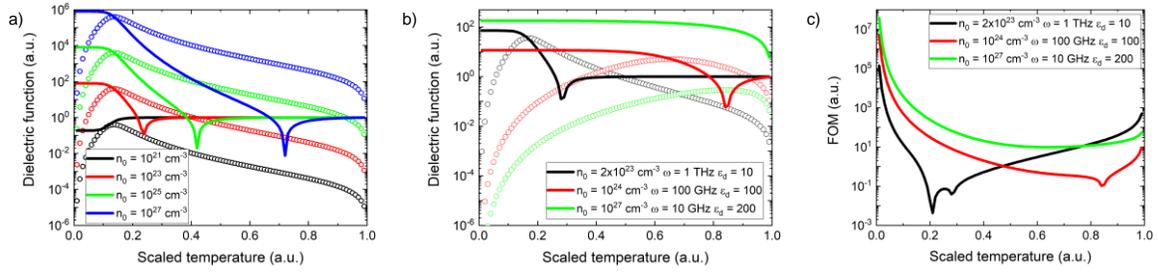

**Figure 3**. (a) Dielectric function of 1D superconductors for varying carrier concentration and temperature. Frequency is fixed to 1 THz. (b) Dielectric function of 1D superconductors for selected combinations of carrier concentration, frequency and surrounding permittivity plotted against changing temperature. (c) Figure of merit results of 1D superconductors calculated for the same parameter combinations as in (b).

Unlike previous description, in this mode the superconducting plasma frequency is bound to the scattering rate. Therefore, the mode could only survive at frequency below the damping frequency instead of all frequencies below the gap frequency.

Fig. 3a shows $\varepsilon_s'$ and $\varepsilon_s''$ change with respect to temperature for different carrier concentrations of quasi-1D superconductor. As can be seen the dielectric function stays constant only for extreme low temperatures. This is because of low permittivity of the surrounding ($\varepsilon_d = 1$) and high plasma frequency (1 THz). This interplay among dielectric and frequency can be solved at various conditions. There are also dips for the propagation constant at certain temperatures, which cause non-propagating, fully absorbed plasma. Fig. 3b demonstrates some possible combinations for stable dielectric function at high temperatures. An interesting phenomenon is relatively high $\varepsilon_s''$ throughout the whole temperature spectrum when compared to 2D and 3D superconductors. When unfolded from Eqn. 10, the imaginary part holds relation to damping frequency, which is a modified version of the plasma frequency. Therefore, there continuously occur an interplay among the real and the imaginary part at high temperatures.

Moreover, this aspect strongly affects the propagation constant and propagation length. However, high ambient permittivity could also potentially lower the imaginary part to extreme values. FOM is plotted in Fig. 3c for the same selected combinations as in Fig. 3b. Similar to previous observations, in all cases there is an exponential drop with temperature. Since the superconductor attains higher number of cooper pairs at low temperatures, the propagation survives for longer periods. On the other hand, the minimum does not occur near $T_c$, which is counter-intuitive. This aspect is a mere reflection of the dips appearing in $\varepsilon_s'$.

## Surface Plasmons on 0D Superconductors

Particles that restrict extensions toward all dimensions are not studied for propagating plasma modes but for localized surface plasmons (LSPs) [20]. For metals, many sensing applications that enable enhancement of incident radiation by means of scattering has been investigated. The 0D case adopts the same equations for plasma frequency and dielectric function from 3D. There are two specific parameters that account for the quality of the LSPs, the scattering and absorption cross sections

$$\sigma_s = \frac{8\pi}{3} k^4 r^6 \left| \frac{\varepsilon_s - \varepsilon_d}{\varepsilon_s + 2\varepsilon_d} \right|^2 \tag{11}$$

$$\sigma_a = 4\pi k r^3 \, \text{Im}\left( \frac{\varepsilon_s - \varepsilon_d}{\varepsilon_s + 2\varepsilon_d} \right) \tag{12}$$

If the real part of the superconductor dielectric function is much larger than the corresponding permittivity of the surrounding ($\varepsilon_d$), the scattering coefficient do not rely on it and simply scale according to the radius. Scattering is more enhanced and overall quality is much better for larger particles. On the other hand, even though absorption does not directly scale with $\varepsilon_s''$, it dies out if $\varepsilon_s''$ is very large or very small compared to $\varepsilon_s'$ and $\varepsilon_d$.

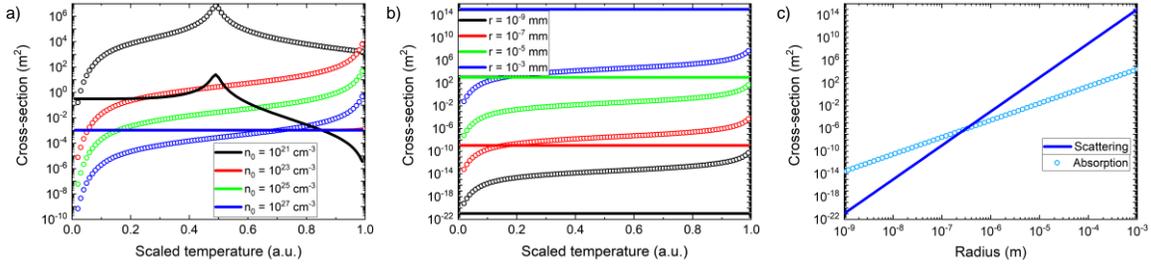

**Figure 4**. (a) Scattering and absorption cross-sections of 0D superconductors for varying carrier concentration and temperature. Radius and frequency are fixed to 1 µm and 1 THz (b) Scattering and absorption cross-sections of 0D superconductors for varying radius and temperature for fixed carrier concentration $n_0 = 10^{28}\ cm^{-3}$. (c) Scattering and absorption cross-sections of 0D superconductors plotted against radius at $t = 0.5$.

Fig. 4a plots the coefficients for varying temperature at different carrier concentrations. The particle radius is set to 1 µm, the incident radiation is 1 THz and vacuum environment ($\varepsilon_d = 1$) is taken. The scattering cross section shows strong temperature dependence only for extreme low concentrations. However, in that case the absorption cross section is superior almost for all temperatures. For higher concentrations, $\sigma_s$ is constant and not temperature dependent as can be expected due to large $\varepsilon_s'$. The absorption, ergo the loss, mostly has linear dependence on temperature except for the exponential behavior at the limits. It is also obvious that highest carrier concentration at low temperatures would be fittest for observation of LSPs. Fig. 4b shows the particle radius variation at a high concentration $n_0 = 10^{28}\ cm^{-3}$. For acute understanding, the sizes are extended from nm to mm. Fig. 4c also gives the direct relation to particle size at a fixed temperature of $t = 0.5$. These two figures explain very well, the applicable range of the LSPs on superconductor particles. Due to gap frequency limiting the radiation, applications relying on small radii become difficult to achieve a lossless plasmonic system.

# CONCLUSION

Plasmonic behavior of superconductors with variety of dimensional aspect are investigated for all possible parameter space. For 3D superconductors, it is found that low carrier concentration is a must for a reliable plasmonic system. Length of a 2D superconductor must be predetermined for any applications since this parameter has direct influence on the final performance. While superconducting nanowires has a radius limit, they are most beneficial if their surrounding ambient has relatively high permittivity. Sub-micron superconducting particles might be problematic since they have large absorption cross-sections, therefore larger particles must be established for long wavelengths. In all of the dimensional cases, temperature is the parameter with the most crucial effect. While low temperature ensures reduced loss, it does not always establish the most ideal plasmonic environment due to lack of surface bound waves.

# COMPETING INTERESTS

The author declares no competing financial interests.

# REFERENCES

1. Homola, J., and Piliarik, M. (2006). Surface plasmon resonance (SPR) sensors. In Surface plasmon resonance based sensors (pp. 45-67). Springer, Berlin, Heidelberg.

2. Zia, R., Schuller, J. A., and Brongersma, M. L. (2006). Near-field characterization of guided polariton propagation and cutoff in surface plasmon waveguides. Physical Review B, 74(16), 165415.

3. Čada, M., and Pištora, J. (2016). Plasmon dispersion at an interface between a dielectric and a conducting medium with moving electrons. IEEE Journal of Quantum Electronics, 52(6), 1-7.

4. Catchpole, K. A., and Polman, A. (2008). Plasmonic solar cells. Optics express, 16(26), 21793-21800.

5. Tsiatmas, A., et. al. (2012). Low-loss terahertz superconducting plasmonics. New Journal of Physics, 14(11), 115006.

6. Khurgin, J. B. (2015). Ultimate limit of field confinement by surface plasmon polaritons. Faraday discussions, 178, 109-122.

7. Majedi, A. H. (2009). Theoretical investigations on THz and optical superconductive surface plasmon interface. IEEE transactions on applied superconductivity, 19(3), 907-910.

8. Carlson, R. V., and Goldman, A. M. (1975). Propagating order-parameter collective modes in superconducting films. Physical Review Letters, 34(1), 11.

9. Buisson, O., Xavier, P., and Richard, J. (1994). Observation of propagating plasma modes in a thin superconducting film. Physical review letters, 73(23), 3153.

10. Tsiatmas, A., et. al. (2010). Superconducting plasmonics and extraordinary transmission. Applied Physics Letters, 97(11), 111106.

11. Tian, Z., et. al. (2010). Terahertz superconducting plasmonic hole array. Optics letters, 35(21), 3586-3588.

12. Thomas, A., et al. Exploring superconductivity under strong coupling with the vacuum electromagnetic field. arXiv preprint arXiv:1911.01459 (2019).

13. Li, M., et. al. (2014). Tunable THz surface plasmon polariton based on a topological insulator/layered superconductor hybrid structure. Physical Review B, 89(23), 235432.

14. Mooij, J. E., and Schön, G. (1985). Propagating plasma mode in thin superconducting filaments. Physical review letters, 55(1), 114.

15. Mirhashem, B., and Ferrell, R. A. (1989). Bias current dependence of the electromagnetic response of thin granular films and Josephson arrays. Physica C: Superconductivity, 161(3), 354-366.

16. Mei, K. K., and Liang, G. C. (1991). Electromagnetics of superconductors. IEEE transactions on microwave theory and techniques, 39(9), 1545-1552.

17. Bennemann, K. H., and Ketterson, J. B. (Eds.). (2008). Superconductivity: Volume 1: Conventional and Unconventional Superconductors Volume 2: Novel Superconductors. Springer Science & Business Media.

18. Maier, S. A. (2007). Plasmonics: fundamentals and applications. Springer Science & Business Media.

19. Cabrera, B., and Peskin, M. E. (1989). Cooper-pair mass. Physical Review B, 39(10), 6425.

20.     Shalaev, V. M., and Kawata, S. (2006). Nanophotonics with surface plasmons. Elsevier.

21.     Gao, F., et. al. (1993). Microwave surface impedance at 10 GHz and quasiparticle scattering in YBa2Cu3O7 films. Applied physics letters, 63(16), 2274-2276.